# Empirical Review of Youth-Employment Programs in Ghana

Monica Lambon-Quayefio

Thomas Yeboah

Nkechi S.Owoo

Marjan Petreski

Catherine Koranchie

Edward Asiedu

Mohammed Zakaria

Ernest Berko

Yaw Nsiah Agyemang

**Abstract**

Ghana's current youth unemployment rate is 19.7%, and the country faces a significant youth unemployment problem. While a range of youth-employment programs have been created over the years, no systematic documentation and evaluation of the impacts of these public initiatives has been undertaken. Clarifying which interventions work would guide policy makers in creating strategies and programs to address the youth-employment challenge. By complementing desk reviews with qualitative data gathered from focus-group discussions and key informant interviews, we observe that most youth-employment programs implemented in Ghana cover a broad spectrum that includes skills training, job placement matching, seed capital, and subsidies. Duplication of initiatives, lack of coordination, and few to non-existent impact evaluations of programs are the main challenges that plague these programs. For better coordination and effective policy making, a more centralized and coordinated system is needed for program design and implementation. Along the same lines, ensuring rigorous evaluation of existing youth-employment programs is necessary to provide empirical evidence of the effectiveness and efficiency of these programs.

**Key words:** youth employment; impact evaluation; Ghana

**JEL Classification**: J08; J21; J48 etc.

**Acknowledgements**

This work was carried out with financial and scientific support from the Partnership for Economic Policy (PEP www.pep-net.org) through funding provided by Mastercard.

## Introduction

Considering the world's youth demographic, youth unemployment and underemployment rates, and the impacts of the COVID-19 pandemic, youth employment has been a priority for governments and development partners and will be for some time to come (Yeboah & Flynn, 2021; International Labour Organization, 2020a, 2020b). Like many other African countries, Ghana faces a significant youth-employment and under-employment problem (Ampadu-Ameyaw et al. 2020). According to the 2021 Population and Housing Census, one in five (19.7%) young persons is unemployed in the country. The unemployment rate for young people between 15 and 24 is even more severe (estimated at 32.8%; see Ghana Statistical Service, 2021). Unemployment is more prevalent among women (15.5%) compared to men (11.6%). Young people are often faced with a unique set of challenges as they transition from school to work, making them more susceptible to unemployment (Oosterom & Yeboah, 2022; Nilsson, 2019). As Avura and Ulzen-Appiah (2016) noted, young people are more likely to remain unemployed as a consequence of to their lack of job-relevant skills, job-search experience, relevant soft and hard skills, and skills mismatch.

In response, the Government of Ghana, through the National Youth Policy enacted in 2010 and its implementation plan from 2014 and 2017, sought to provide guidelines for the implementation of youth-employment programs (hereafter, YEP). The youth policy was recently revised with the overarching theme of "Benefit for Youth and Involve Youth: Together for a Prosperous Future" (Ministry of Youth and Sports, 2022). The revised policy places a high degree of importance on youth participation in the design and implementation of programs and interventions designed for young people. Over the years, several interventions have been designed and implemented by successive governments, concerns in the private sector, nongovernmental agencies, and international and civil-society organizations to deal with the rising unemployment problem among the youth.

In a World Bank inventory exercise, Avura and Ulzen-Appiah (2016) reported more than forty youth-employment programs in Ghana, out of which eight were public-sector led and twenty-two were private-sector led. These initiatives largely focused on skills development and training, job placement, apprenticeship, entrepreneurship training, job search services, and direct employment (Dadzie, Fumey & Namara, 2020; Avura & Ulzen-Appiah, 2016; and Yeboah & Flynn, 2021).

While the government and the private sector have shown determination in resolving issues related to youth unemployment, notable challenges have hindered progress toward the achievement of broad objectives. In addition to fragmentation and lack of coordination in the implementation of youth-employment initiatives, duplication of programs and shifts in program prioritization, beneficiary targeting, and implementation of programs have been the bane of most of these public initiatives (Sumberg et al., 2021; Dadzie, Fumey & Namara, 2020; Kluve et al., 2019; Fox & Kaul, 2017; Fox & Gandhi, 2021). Because the number of new labour-market entrants outweighs job creation, and technological changes limit job prospects, Ghana would need to take urgent and concerted action to realize the government's goal of addressing youth-employment challenges.

The lack of systematic documentation and evaluation of the outcomes of youth-employment programs is a crucial problem as well (Kluve et al., 2019; (Yeboah & Flynn, 2021). The absence of rigorous impact evaluations of these initiatives over the years has made it difficult for policymakers and development partners to understand what interventions work best to improve youth-employment outcomes in Ghana.

In particular, a comprehensive understanding of public youth-employment programs, which is required to guide program design and to make modifications that accurately reflect the challenges implementers and beneficiaries face, is limited. For instance, Avura and Ulzen-Appiah (2016) and Dadzie, Fumey, and Namara (2020) focused on a broad range of public and private programs but only provided brief descriptions of the programs without discussing gaps associated with program implementation. Previous studies have also not provided information that has captured the perspectives of both implementers and beneficiaries. While Dadzie, Fumey, and Namara (2020) attempted to engage stakeholders in their data collection, for example, they focused largely on ministries, agencies, and youth organizations, though these stakeholders were arguably not in a position to provide more disaggregated information on a program-by-program basis as required. For example, youth organizations that may not have been in a position to echo the actual experiences of beneficiaries of specific YEP were consulted rather than individual beneficiaries, nor were all public-program stakeholders consulted (e.g., the Nation Builders Corps, hereafter, NABCO). As such, the information available in the literature is general and does not reflect the nuanced perspectives of implementers and beneficiaries. These are the gaps that our work seeks to fill.

By focusing only on public programs, we were able to delve deeper to provide a detailed and comprehensive understanding of government-funded youth-employment programs by documenting challenges from the perspectives of the major categories of stakeholders, including beneficiaries. Specifically, in addition to the systematic desk review, we provide evidence regarding program effectiveness and gaps not only from the view of government agencies responsible for designing the programs at higher institutional levels, but also from implementers at the district levels who are responsible for day-to-day implementation. Such a detailed understanding is critical to providing policy makers with insights that can inform decisions regarding program modification/re-design or strengthen existing programs.

## Methods

We used a desk review of policy, legislative instruments, and documents to select seven main programs out of the total of twenty-three that have been implemented since the year 2000 to address issues of youth unemployment and underemployment in Ghana. In addition, primary data were collected using semi-structured interviews and focus-group discussions with stakeholders who were identified through a rigorous stakeholder-engagement strategy. For each YEP, in-depth interviews with program implementers and key officials from relevant government agencies were conducted. Focus-group discussions were also conducted with beneficiaries of the selected YEP.

The main criterion for selecting YEP for inclusion in this evidence review was that the program fell under either public or public-private partnership. A total of twelve public YEPs remained after all exclusions. We subsequently focused on seven, excluding five that never implemented, had yet to take off, or were at the early stages of implementation at the time of our research. The selected programs capture a variety of target groups within the larger youth population: while some targeted youth with a post-secondary education (e.g., the Nation Builders Corps, hereafter NABCO), others targeted semi-literate youth (e.g., the National Entrepreneurship and Innovation Program, hereafter, NEIP). The YEP we selected also covered a variety of program types, including skills training with elements of subsidies and job placement, reflecting a wide-range of intervention areas for youth employment (skills training, job-matching, entrepreneurship, job creation).

The selected programs include the Youth-Employment Module of the Youth Employment Agency (hereafter, YEA), NABCO, Youth in Afforestation, Youth in Agriculture, the National Entrepreneurship and Innovation Program, the National Service Scheme, and the Rural Enterprises Program (Youth-in-Agribusiness Component). Taken together, these programs represent some of the largest youth-employment initiatives that Ghana has implemented since the turn of the millennium. Building on the work of Avura and Ulzen-Appiah (2016) and of Dadzie, Fumey, and Namara (2020), we combined in-depth review of policy documents with focus-group discussions and in-depth interviews of individual stakeholders through a purposive sampling technique. Our in-depth interviews and focus-group discussions provided a more nuanced understanding of the progress and challenges of youth-employment programs, including lapses in program design, gaps in implementation, and the experiences of beneficiaries of the YEP.

In all, a total of fifty-two individual interviews and twelve separate focus-group discussions were conducted across the seven selected YEP in four regions (Greater Accra, Ashanti, Bono, and Northern regions) that reflect Ghana's ecological zones (see Table 1). Out of fifty-two in-depth interviews participants, eight were key informants, including the directors of programs or institutions responsible for implementation. To ensure that we captured the perspectives of youth across the country, stakeholder engagements were undertaken in the four study regions. These regions were chosen because of the population of youth (i.e., Greater-Accra and Ashanti) resident there and because most of the agriculturally-related YEP targeted the Bono and Northern regions. Two validation workshops were conducted in the Ashanti and Greater-Accra regions to validate the findings from the study.

Table 1: Summary of Field Work by Program

| Program | Key Informant | IDI (Beneficiaries) | FDG |
| --- | --- | --- | --- |
| National Service Scheme (NSS) | 1 | 10 | 3 |
| Nation Builders Corps (NABCO) | - | 2 | 3 |
| Youth in Afforestation Program (YAFP) | - | 9 | 2 |
| Youth in Agriculture (YIA) | 1 | 1 | - |
| National Entrepreneurship and Innovation Program (NEIP) | 1 | 3 | 1 |
| Rural Enterprises Program (REP) | 1 | 4 | 1 |
| YEP, under the Youth Employment Agency (YEA) | 2 | 15 | 2 |
| TOTAL | 6 | 44 | 12 |

In addition, we supplemented qualitative data with information from budget statements from the Ministry of Finance. We also conducted in-depth desk reviews of relevant government reports and policy documents, including the annual performance reports of the ministries and agencies that are responsible for implementing the selected YEP. In addition, we reviewed published journal articles and technical and working papers, especially those from the World Bank that have examined the impacts and identified gaps in YEP. In instances in which we lacked information regarding specific programs, we augmented our data with in-depth interviews and focus-group discussions with relevant stakeholders. Both the Ghana Living Standards Survey (Ghana Statistical Service, 2018) and the Ghana Demographic and Health Survey are nationally representative data and therefore provide descriptive statistics on the labour market, education outcomes, and demographic characteristics representative of youth in Ghana. Recent unemployment rates and other labour-market outcomes were obtained from the 2021 population and housing census report.

**Findings**

*Review of Youth-Employment Programs*

All the YEP in our study are public programs implemented in various sectors of the economy and have a significant if not exclusive focus on youth. Each program is also specifically focused on a type of youth—e.g., those within a specific age bracket or location or with specific skills set.

*Youth Employment Module of the Youth Employment Agency*

Ghana's Youth Employment Agency is a public-sector institution that runs the youth-employment modules intended to create job opportunities for young people who are willing and able to work. The modules cover a wide range of sectors including agriculture, community service, education, health, sanitation, trade and vocation, ICT, and sports (Dadzie, Fumey & Namara, 2020). Additional modules have been developed and implemented in line with the services that may be required by the country at a particular period. The YEP modules target youth who do not have tertiary education aged 15-35, including minority groups and persons with disabilities, who are recruited for and placed in job opportunities to enable them gain entrepreneurial training, employable skills, and employment services to facilitate their transition into the labor market (Ampadu-Ameyaw et al., 2020). Selection of beneficiaries is facilitated by YEA officials through regional and district offices across the country. Each module has its own panel of experts constituted to support the selection of beneficiaries. Although the YEA has implemented all planned modules and has reached many young people in low-income households and those out of school, the program has its weaknesses, including limited and irregular government funding, inability to offer long-term wage employment, absence of an exit strategy for beneficiaries, weak monitoring and evaluation systems (Avura & Ulzen-Appiah, 2016; Dadzie Fumey & Namara, 2020), and local operational challenges such as the absence of properly constituted committees for selecting qualified youth.

*Nation Builders Corps (NABCO)*

The Nation Builders Corps is a skills-training program launched in 2018 to provide temporary employment to and improve the employable skills of young graduates in the public sector (Nation Builders Corps, 2017). The program is open to all tertiary level graduates between the ages of 18 to 35 who have completed national service and are unemployed, regardless of vulnerability or disability (Dadzie, Fumey & Namara, 2020; Atiemo, Quaynor & Asante, 2020). The program covers sectors such as agriculture, education, health, revenue mobilization, sanitation and governance, digitization, and private-sector support (Dadzie, Fumey & Namara, 2020; Atiemo, Quaynor & Asante, 2020). Selection is done by the NABCO secretariat, and beneficiaries receive an allowance of GhC 700 (USD $80.28) every month for a period of three years. It is funded by the government and is monitored and evaluated by the NABCO secretariat through quarterly reports. The NABCO program has provided opportunities for 100,000 young graduates, with 33,000 reportedly obtaining employment after graduation., according to Modern Ghana media report (2021)2021, The program, which was set to end in October 2021, was extended for another year, and the government plans to absorb remaining beneficiaries into a new program called YouStart. While no data is available on the percentage age of beneficiaries who have or secured employment, beneficiaries cited training as the most important outcome, and the main problem reported was a delay in receiving the monthly support/allowance. Despite this, beneficiaries were generally happy to have a job, and the online recruitment process was considered one of the program's main strengths.

*Youth in Afforestation Program (YAFP)*

The Youth in Afforestation program is a government initiative aimed at creating employment for young people by increasing demand for their labour in tree planting to restore degraded lands caused by illegal small-scale mining (Ampadu-Ameyaw et al., 2020); Government of Ghana, 2018). The program targets young people aged 18-35 with a minimum qualification of Higher National Diploma (HND) and good communication and organizational skills. Beneficiaries receive a two-year contract with a monthly allowance of GhC 700 and The program is ongoing and funded through government allocations to the forestry sector and is monitored and evaluated through quarterly reports at the district level. As of 2019, approximately 65,000 youth had benefited from the program, and around twenty-six million seedlings of various tree species had been planted. An evaluation by Nketia et al. (2022), however, revealed that the program faced various challenges, including financial, logistical, and political hurdles that reduced the program's effectiveness in transforming Ghana's socioeconomic and environmental landscape. Despite this, the program has promoted landscape diversity in former farmland areas and has increased public awareness of forest protection (Nketia et al., 2022).

*National Entrepreneurship and Innovation Program (NEIP)*

The National Entrepreneurship and Innovation Program is a government-funded program that provides training, incubation, funding, and policy direction to young entrepreneurs aged 18-35 in various sectors including media, agribusiness, fashion, and Information and Communication Technologies (ICT). The program selects beneficiaries through a two-stage approach. First all applicants receive training and business advisory services and, in the second stage, beneficiaries who demonstrate growth potential are recommended for funding. The NEIP's goals are to enhance industrial growth, create jobs for youth, and achieve Ghana's long-term vision of consolidating its middle-income status. The program is inclusive, targeting young people irrespective of gender, social class, or disability status, and has a web-based application portal (NEIP, 2022). The NEIP received a total government investment of $155 million, with a budgetary allocation of GhC 47

million in 2019 (Ampadu-Ameyaw et al., 2020); Armah, 2018). So far, more than 45,000 entrepreneurs have been trained, and over 9,000 have received funding. There has never been a rigorous evaluation of the program's outcomes, however. The NEIP has launched new initiatives, including a business competition to empower women in technical and vocational education and training trades and the establishment of 1,000 greenhouse projects to create jobs.

*Youth in Agriculture Program*

The Youth in Agriculture program, introduced in 2009, was intended to create employment opportunities for young people in the agricultural sector and improve their standard of living (Ampadu-Ameyaw et al., 2020); Ohene, 2013; Ministry of Food and Agriculture (2011). The program offers young farmers technical support, training, material credit, inputs, and support to acquire large hectares of land for farming. The program has four main components: crop/block farm, livestock and poultry, agribusiness, and fishery and aquaculture and targets all young people in rural areas, aged 15-35, who are interested in pursuing a career in the agricultural sector. The program is implemented by the Ministry of Food and Agriculture and funded by the government with an estimated amount of GhC 42 million (USD $4.9 million) allocated in 2018 (Ampadu-Ameyaw et al., 2020). Although it is unclear how many people have benefited from the program since its inception, the number of annual beneficiaries reportedly increased from 50,000 in 2016 to approximately 80,000 in 2020 (Avura & Ulzen-Appiah, 2016; Ampadu-Ameyaw et al., 2020). A subset of the program in partnership with an agrochemical company has shown promising results, with an increase in acres of land under cultivation for maize and rice production. Youth in the program face challenges, however, including a lack of awareness of the incentive packages and benefits and difficulties in securing land ownership (Ohene, 2013).

*National Service Scheme*

The Ghana National Service Scheme (hereafter, NSS) was established in 1973 to provide newly qualified graduates with practical, on-the-job experience in both the public and private sectors, while instilling a sense of national pride in participants. The scheme offers short-term employment services and entrepreneurial training through community service and covers all sixteen administrative regions of Ghana. Graduates of accredited public and private tertiary and professional institutions, including young men and women aged 18 and above, can apply, with no interviews before placement. NSS personnel receive an allowance of GhC 559.04 per month. Forty percent of the scheme is funded by national budget allocation, 40% by internally generated funds, and 20% by private-sector user organizations and agencies Dadzie, Fumey & Namara, 2020). It is supervised and monitored by the Ghana Education Service and National Service Secretariat. The monitoring and evaluation mechanisms of the NSS program are inadequate, with no tracer studies or large-scale evaluations. Approximately 70,000 tertiary graduates are assigned by the NSS each year. No rigorous impact evaluation of the program has ever been undertaken, however. Recently, the Abdul Latif Jameel Poverty Action Lab attempted to evaluate components of the program by randomizing and examining migration, assimilation, and labour-market outcomes. Results from this evaluation are not yet available. Audit reports suggest that service personnel were not properly oriented before being assigned to their posting and were not adequately monitored while on the job.

*Youth-in-Agribusiness Component of the Rural Enterprise Program*

The Rural Enterprise Program (hereafter, REP) is a government initiative aimed at reducing rural poverty and creating employment opportunities through the establishment of micro- and small-scale enterprises. The program's three main components are access to business development services, technology transfer through training and demonstrations, and access to rural finance through participating financial institutions. The Youth-in-Agribusiness component targets young people between the ages of 18-35 with an interest in pursuing agribusiness and provides them with capacity building training, agro-processing, farm-based start-up kits, and pre-financing support. Beneficiaries may receive a matching grant and cash from the Rural Enterprises Development Fund after completing the training. The program operates in all sixteen administrative regions of Ghana and is evaluated through tracer studies, outcome surveys, and performance evaluations. Various institutional co-financing arrangements fund the program for a total of $249.87 million, although specific funding for the Youth-in-Agribusiness Component was not provided by key informants.

The REP, which includes the Youth-in-Agribusiness Component, has been running since 2011 and is expected to end in 2024, but data on the number of youth beneficiaries is limited.

## Political Economy of Youth-Employment Programs

*The (In)Adequacy of Youth-Employment Programs*

The reviewed YEP have shown potential in providing dignified and fulfilling work for youth and for vulnerable groups, including women and people with disabilities. It appears that most programs have achieved some aspects of their objectives to an extent but not without challenges. The lack of standards in the measurement of the actual impact of the YEP make it difficult to ascertain the effectiveness of the programs in achieving their objectives. Beyond the data that is reported in annual progress reports on the number of beneficiaries, there have been no rigorous impact evaluations of the programs to provide robust empirical evidence on their effectiveness.

YEP are dominated by skills training and direct job placement although they offer only temporary employment for beneficiaries. Most YEP can be considered largely inclusive because women and individuals with disabilities have not been excluded. In fact, in some instances (e.g., the NEIP), provisions have been made for individuals with disabilities even when they did not satisfy the selection criteria.

For all the YEP considered in this study, however, no specific quotas for persons with disabilities were indicated, making it difficult to make a firm conclusion that vulnerable groups have adequately been served under YEP in Ghana. Stakeholders only indicate by word of mouth that the program considers people who are vulnerable although this is not explicitly documented.

Steps have been taken by the mangers of the YEPs to ensure that there is minimal interference by the government to ensure effective implementation of these YEPs in Ghana. This may perhaps be the result of the unfavorable perception the public holds regarding widespread lack of transparency about youth-employment programs in the recent past. For instance, to eliminate misconduct and ensure transparency in procurement, institutional restructuring of the Youth in Agriculture Program resulted in the transfer of responsibility for procurement of agricultural inputs from the directors to private commercial agents who were expected to provide inputs at competitive rates. However, not all programs were able to avoid government interference totally as stakeholders report some degree of interference by government officials in the implementation of programs.

Our review showed a great deal duplication and overlap, consistent with the conclusions of Dadzie, Fumey, and Namara (2020), who reported duplication of effort as a key factor that affected youth-employment programs. The duplication of programs coupled with the lack of coordination creates a situation in which limited resources may not be put to optimal use. For example, we found that similar "youth in agriculture" modules were being implemented by the Ministry of Food and Agriculture and by the Youth Employment Agency. Similarly, we also found that entrepreneurship programs were being rolled out independently by the NEIP and the NSS without proper structures in place to coordinate efforts.

Policymaking regarding youth-employment issues in Ghana is convoluted, and a total of nine ministries, two coordinating agencies, and several other public institutions all work on various programs related to youth employment. While the Ministry of Employment and Labour Relations is mandated to coordinate stakeholder activities related to youth-employment programs, their role is often limited when programs are launched directly under the office of the president. Example of such programs are NABCO and the NEIP programs which are directly supervised at a special desk in the office of the president. Direct control from this office constrains the Ministry's ability to carry out supervisory and coordinating roles relating to the design and implementation of the programs. This view reflects the concerns shared by some key informants who indicated that government ministries or agencies always have trouble exercising oversight over coordination, supervision, or monitoring and evaluation of special flagship programs initiated and implemented by special desks at the office of the president.

Moreover, our analysis demonstrated that selection processes for most of the YEP studied were characterized by external influence by political elites, traditional leaders, and other influential individuals, which limits the participation in YEP of young people with few or no social connections with elites or politicians. In some instances, applicants may not meet eligibility criteria or even be interested in participating in the YEP but are able to gain employment because of the influence of chiefs and politicians they may be associated with. A key informant noted:

> *Sometimes you'll just be there and the Omanhene (Chief) will come and say* to the *director,* "This is *my person.* Please *help him."* Meanwhile*, the person does not even have passion for the work. You get it? The way of recruitment, the external influence, from powerful people in society is not easy. If today as Metropolitan Director we are doing recruitment,* an important chief *can walk in and say*, "This *is my nephew or niece."* Then *I look at him and say*, "It's *not possible?" Who am I to say otherwise? Do you understand?*

The keen interest of politicians, chiefs, and other influential persons to ensure that their own people become beneficiaries of Youth Employment Agency is related to their desire to be seen as having succeeded in securing employment for their people. This is summed up perfectly in the words of one official from a YEP:

> *They only show interest when there's recruitment. So that they will bring in their people, that is when they show interest. All of them. The Omanhene [Chief] wants to feel that he has been able to get employment for his people, and the politician will also feel that as a politician … I can use it as part of my campaign message of the number of jobs I have gotten for people.*

While this may happen in some cases, it does not overshadow the objectivity of the recruitment process for all YEP. Key informants indicated that, despite external influences, all applicants, whether supported or not by an influential person, must still meet the entry criteria and go through the selection process. Similarly, some beneficiaries of the NABCO program intimated that the selection process for NABCO was not completely because given that some applicants who met the criteria and applied for the program never got selected because of their lack of social connections. However, one young man interviewed explained that the process and selection of beneficiaries for the NABCO program was fair and transparent:

> *I would say something about NABCO. I did some and if I say that I encountered the issues of protocol, I would be telling lies. I just applied; I didn't have to see anyone. I used the online portal as they taught us. I typed in my particulars, and they called me for an interview. It is the only interview that I did not have to know anyone to get it. Everyone who was selected was selected based on their certificates because all those selected for my department had the same certificate. The application process was not hard. Even if you're not conversant with online applications, you can go to a café and get help.*

We also found that, while the standard procedure for the Youth in Afforestation Program involved applying through a secured online portal or with the YEA district offices, young beneficiaries revealed that the selection process was not inclusive. According to them, the process was characterized by a complex web of favoritism to the extent that several beneficiaries on the program had not undergone the normal application process.

By far, our analysis highlights that the sustainability of YEP in Ghana is under threat. This is mainly because, for most of the YEP, sustainability is not taken into consideration at the design stage, which is perhaps because many YEP are political tools designed and implemented by government institutions to achieve a particular objective. As noted by a key informant at the Ministry of Employment and Labour Relations, the lack of extensive engagement and specific strategies to graduate beneficiaries from one phase of the program to another threatens viability. For example, the lack of an exit strategy for NABCO created a situation whereby most beneficiaries were still unemployed after participating in the program for three years.

Moreover, limited budgetary allocations have made it difficult for some components of YEP to be sustained. A key informant from the NEIP disclosed that the lack of budgetary allocations made it difficult for the beneficiaries to receive support in the second year of the program. Similarly, interviews with the National Youth Authority revealed that limited budgetary allocations significantly affected their operations. Aside from monetary constraints, issues of limited administrative capacity were also raised. Though YEP are required to operate in all districts, the lack of personnel restricts their activities of some programs to only a few areas of the country. In particular, concerns have been raised regarding the limited presence across the country of the National Youth Authority and the NEIP because of a lack of administrative capacity. The National Youth Authority has been forced to rely on temporary staff (e.g., national service personnel) for their activities in some districts.

*Endogenizing Youth-Employment Programs*

As noted earlier, the design and implementation of many of the youth-employment programs in Ghana are political tools which various government and politicians have used to achieve their objectives. As such, the design process is often short-sighted and inputs from government agencies mandated to coordinate and oversee such programs are limited. Most often, the design of youth-employment programs is linked to the political process. Rather than continuing existing programs, successive governments prefer to design and implement new programs similar in structure and scope either to appear relevant or to be used as a tool to win elections. For instance, the establishment of NABCO was specifically designed to address rising levels of unemployment among graduates who had, over the years, mobilized to become a powerful force in communicating and advocating for their members. The emphasis on job creation for the youth in the 2020 elections, for instance, contributed to a situation where new but duplicative entrepreneurship programs such as the NEIP were established. The administrative structures of such programs, where the main administration and implementation of the program is controlled from the presidency, restrict the mandates of existing government structures such as the ministry of employment and labour relations to exercise their oversight roles which has implications for resource distribution of such programs.

Also, in some instances, after promises are made to youth groups regarding the creation of special youth programs to solve specific aspects of the youth-employment problem, lack of funds makes it difficult for the programs to be implemented. In 2011, for example, the graduate business-support scheme, a private-public partnership was launched to provide business skills training to support graduates across the country in starting their own businesses. Although the target was to organize about 500 business clinics within a year to provide business skills, the program suffered due to the government's inability to secure funds for implementation. As a result, the program remained dormant and was never revived by any successive governments.

With most of the YEP designed as political tools to execute an agenda that elevates a particular government in power, the directors of the YEP are usually politically aligned. The limitation of this is the fact that such political appointments encourage rent-seeking behaviour with less transparency as has been witnessed in such past programs as the Ghana Youth Employment and Development Agency in which lack of transparency, among other factors, led to the eventual collapse of the agency.

The role of the private sector in youth-employment-policy making and design is limited. Although there has been some active participation by non-governmental organizations, civil-society organizations, and some international partners, the government dictates the design and implementation of YEP to a substantial extent. In a few instances, however, international organizations such as the Mastercard Foundation may partner with the government to design and implement initiatives for reducing youth unemployment.

### Conclusions and Policy Implications

In dealing with the youth-employment challenge in Ghana, various governments have dedicated significant resources in designing and implementing interventions in various sectors of the economy. Despite huge investment in these programs, there is little to no empirical evidence on the effectiveness of youth-employment programs in Ghana. This study attempted to review public youth-employment programs that have been implemented over the years by documenting programmatic gaps, obstacles to implementation, and the political-economy issues that have plagued the design and implementation of these YEP. In addition, we hope to have provided a more nuanced understanding of these issues by complementing secondary data and desk reviews with qualitative interviews with a range of relevant stakeholders.

We found that the YEP implemented in Ghana range from skills training to job placement matching, seed capital, and subsidies. While some programs focus on one type of support (e.g., skills training or job placement), other YEP combine two or more types of support (such as NEIP's combination of entrepreneurship training and seed capital) or a combination of skills training and subsidies as in the Youth in Agriculture Program. Youth in all regions in the country are targeted beneficiaries. However, the nature of some program such as youth in agriculture types restricts them to specific sectors. Moreover, while YEP caters generally to young people between 15 and 35, some programs have additional specifications such as youth found in the rural areas, tertiary and professional graduates, senior secondary school graduates, dropouts etc. At the same time, none of the YEP we considered specifically target or provide quotas for youth groups that are considered vulnerable or with disabilities.

Although some YEP follow the selection process as indicated in the design phase, for some of the programs, the selection of beneficiaries is not transparent. Interviews with beneficiaries and program directors suggest that, in some cases, selection is influenced by politicians, traditional leaders, and people with power or political connections resulting in a lack of fairness in the implementation of the programs. We also found several duplications of interventions by YEP, which may be associated with the politicization of design and implementation. In view of the sensitive nature of youth unemployment in Ghana, governments have often used YEP as political tool to remain popular or to garner votes during elections, eliminating the incentive to strengthen existing YEP. In most cases, there is no coordination among implementing agencies even when they provide similar support. Also, we found that almost all the YEP have not undergone any robust impact evaluation to ascertain whether these programs are yielding the desired impacts on youth unemployment. Indeed, only a few of the programs produce annual reports that are available to the public.

These findings have important implications for policy making regarding the design and implementation of YEP in Ghana. Centralization of YEP under one ministry/agency could minimize program duplication and strengthen coordination among various youth-related agencies and units of government. In addition, consideration of exit strategies for beneficiaries at early stages of YEP design would ensure the creation of long-lasting jobs for youth. At the same time, institutionalizing impact evaluations of YEP and using the findings from those evaluations would ensure efficient resource use and improved design and implementation of future YEP. The government could also consider starting with sector-based harmonized youth programs which consolidate all YEP at the sector level along with evaluations to ascertain the impact of the YEP. The implementation of sector-based harmonization could serve as a steppingstone to centralizing the design and implementation of all YEP under one ministry.

The findings from this study suggest directions for future work. First, given the lack of robust impact studies, evaluation using quasi-experimental techniques for existing programs or randomized controlled trials for programs that have yet to be implemented would be crucial. In such studies, both short-term outcomes and longer-term impacts should be considered. Second, our work revealed that many YEP lack exit strategies which makes them less effective in dealing sustainably with the problem of youth unemployment. We currently have limited knowledge about exit strategies that are effective in Ghana's context. Future research should, therefore, provide a deeper understanding of exit strategies and should determine their cost effectiveness. Third, while it is important to capture the benefits or impacts of YEP on individual beneficiaries, it is not clear from the literature what the broader economy-wide impacts of such programs are. It would be useful to understand the macro-level impacts of YEP to guide policymaking and resource allocation.